\documentclass[numberedappendix]{emulateapj}
\begin{document}

\slugcomment{Accepted for publication in ApJ: January 29, 2008}

\title{Gas and Dust Emission at the Outer Edge of Protoplanetary Disks}

\author{A. M. Hughes \altaffilmark{1}, 
D. J. Wilner \altaffilmark{1}, 
C. Qi \altaffilmark{1}, 
M. R. Hogerheijde \altaffilmark{2}
}

\email{mhughes@cfa.harvard.edu}

\altaffiltext{1}{Harvard-Smithsonian Center for Astrophysics,
  60 Garden Street, Cambridge, MA 02138}
\altaffiltext{2}{Leiden Observatory, Leiden University, P.O. Box 9513, 2300 RA,
  Leiden, The Netherlands}

\bibliographystyle{apj}

\begin{abstract}
We investigate the apparent discrepancy between gas and dust outer 
radii derived from millimeter observations of protoplanetary disks.  
Using 230 and 345~GHz continuum and CO J=3-2 data from the 
Submillimeter Array for four nearby disk systems (HD 163296, TW Hydrae, 
GM Aurigae, and MWC 480), we examine models of circumstellar disk 
structure and the effects of their treatment of the outer disk edge.
We show that for these disks, models described by power laws in surface 
density and temperature that are truncated at an outer radius are 
incapable of reproducing both the gas and dust emission simultaneously: 
the outer radius derived from the dust continuum emission is always 
significantly smaller than the extent of the molecular gas disk traced by 
CO emission.  However, a simple model motivated by similarity solutions of the 
time evolution of accretion disks that includes a tapered exponential edge in 
the surface density distribution (and the same number of free parameters) does 
much better at reproducing both the gas and dust emission.
While this analysis does not rule out the disparate radii implied by the 
truncated power-law models, a realistic alternative disk model, grounded 
in the physics of accretion, provides a consistent picture for the extent
of both the gas and dust. 
\end{abstract}


\section{Introduction}

Characterizing the gas and dust distribution in the disks around young stars
is important for understanding the planet formation process, as these disks
provide the reservoirs of raw material for nascent planetary systems.  A common 
method of modeling circumstellar disk structure is to use models described by 
power laws in surface density and temperature that are truncated at
a particular outer radius.  This prescription has its historical roots in
calculations of the minimum mass solar nebula, which indicated a surface
density profile of $\Sigma \propto r^{-3/2}$ \citep[e.g.][]{wei77}, 
as well as theoretical predictions of a radial power-law dependence of
temperature for accreting disks around young stars \citep{ada86,ada87}. 
Observationally, the parameterization of temperature and surface density as 
power-law functions of radius began with early spatially unresolved studies of 
continuum emission from disks \citep{bec90,bec91}.  These models have since 
been refined and applied to spatially resolved observations of many disks with 
success \citep[e.g.][]{mun93,dut94,lay94,dut98}, and they have proven useful 
for understanding the basic global properties of disk structure.  Recently, 
however, with the advent of high signal-to-noise, multi-frequency observations 
of gas and dust in protoplanetary disks, these models have begun to encounter 
difficulties, particularly in the treatment of the outer disk edge.

The extent of the gas and dust distribution in circumstellar disks has 
implications for our understanding of the planet formation process in our
own solar system.  There is some evidence for a sharp decrease in the surface 
density of Kuiper Belt objects beyond a distance of 50 AU from the Sun 
\citep{jew98,tru01,pet06}.  However, the origin of this edge is unclear.  
\citet{ada04} note that the observed distance is far interior to the 
radius at which truncation by photoevaporation would be expected to occur, 
while \citet{you02} find that the presence of such an edge in planetesimal 
density could be explained by drift-induced enhancement.  A compelling
possibility is that the Sun formed in a cluster environment, and the early 
solar disk was truncated by a close encounter with a passing star 
\citep[see][and references therein]{rei05}.  A more complete
understanding of the outer regions of protoplanetary disks may provide 
insight into the processes that shape the outer solar system.  

\citet{pie05} present multiwavelength millimeter continuum and CO isotopologue 
observations of the disk around the Herbig Ae star AB Aurigae and found from 
fitting models of disk structure described by truncated power laws that the 
outer radius of the dust derived from continuum emission ($350\pm30$~AU) was 
much smaller than that of the gas derived from $^{12}$CO J=2-1 emission 
($1050\pm10$~AU).  They suggest that a change in dust grain properties 
resulting in a drop in opacity could be responsible for the difference, and 
note the possible association with a ring feature in the disk at 200~AU.  
A similar result was obtained by \citet{ise07} from observations of the disk
around the Herbig Ae star HD 163296: they found a significant discrepancy 
between the outer radius derived for the dust continuum emission 
($200 \pm 15$ AU) and that derived from CO emission ($540 \pm 40$ AU).  
These data appeared to require a sharp drop in surface density, opacity, 
or dust-to-gas ratio beyond 200~AU; however, as they discuss, there is no 
obvious physical basis for such a discontinuity.  As \citet{ise07} 
demonstrate, the discrepancy in outer radii derived from the dust and gas
is not simply an issue of sensitivity; the observations were sufficiently 
sensitive to detect emission from the power-law dust disk if it did extend 
to the radius indicated by the CO emission.  The underlying issue is that the 
truncated power law model does not simultaneously reproduce the extent of 
both the continuum and CO emission for these disks.  

Using data from the Submillimeter Array\footnote{The Submillimeter Array is a 
joint project between the Smithsonian Astrophysical Observatory and the 
Academia Sinica Institute of Astronomy and Astrophysics and is funded by the 
Smithsonian Institution and the Academia Sinica.}
we show that the same apparent discrepancy in gas and dust outer radius
applies to the circumstellar disks around several more young stars.
In an attempt to understand the origin of this discrepancy, we investigate an 
alternative surface density profile based on work by \citet{har98}, which is 
similar to a power law profile in the inner disk but includes a tapered outer 
edge.  We show that this model, which has a physical basis in similarity 
solutions of disk evolution with time, is capable of simultaneously 
reproducing both continuum and CO emission from these disks.
The primary difference between 
this model and the truncated power-law disk is that instead of a sharp outer 
edge the surface density falls off gradually, with sufficient column density
at large radii that CO emission extends beyond the point at which 
dust continuum emission becomes negligible.  

\section{Millimeter/Submillimeter Dust Continuum and CO J=3-2 Data}

The analysis was conducted on extant SMA data of the disks around
of HD 163296, TW Hydrae, GM Aurigae, and MWC 480.  
The dates, frequencies,
antenna configurations, number of antennas, and original publications 
associated with the data sets are listed in Table \ref{tab:obs}.
The four disk systems chosen for this analysis are all nearby, bright, 
isolated, and have been well studied at a wide range of wavelengths.  
The velocity fields of these disks all appear to be well described by 
Keplerian rotation \citep{ise07,qi04,dut98,pie07}.  The relevant properties 
of these systems (spectral type, distance, stellar mass, age, and 
disk inclination and position angle) are listed in Table~\ref{tab:params}.  

\begin{table*}
\centering
\caption{Sources of SMA 230/345 GHz continuum and CO J=3-2 data.}
\begin{tabular}{clcccc}
\hline
Freq./ & & & Array & No. of & \\
Transition & Object & Dates & Config. & Antennas & Reference\\
\hline
\hline
230 GHz & HD 163296 & 23/24 Aug 2003 & Compact N & 7 & 1 \\
 & TW Hydrae & 10 Apr 2005 & Extended & 8 & 2 \\
 &        & 27 Feb 2005 & Compact & 8 & 2 \\
 & GM Aurigae & 10 Dec 2006 & Extended & 8 & 3 \\
 & MWC 480 & 18/20 Nov 2003 & Compact N & 8 & 1 \\
\hline
345 GHz/ & HD 163296 & 23 Aug 2005 & Compact & 8 & 4 \\
CO J=3-2 & TW Hydrae & 28 Dec 2006 & Compact N & 8 & 5 \\
 & GM Aurigae & 26 Nov 2005 & Compact & 7 & 6 \\
 & MWC 480 & 21 Oct 2005 & Compact & 8 & 1 \\
\hline
\end{tabular}
\tablerefs{~
(1) SMA archive; 
(2) \citet{qi06}; 
(3) Qi et al. (in prep); 
(4) \citet{ise07} ; 
(5) Qi et al. (2007, submitted); 
(6) \citet{and07beta}.
}
\label{tab:obs}
\end{table*}

\section{Disk Models}

Using the SMA data available for the four disk systems, we compared two 
classes of disk models: the first model is described by power laws in surface 
density and temperature and is truncated at an outer radius $R_{out}$ 
(details in \S\ref{sec:pl}), and the second model is described by a 
power law in temperature and a surface density profile similar to a power 
law in the inner disk but tapered with an exponential edge in the outer 
disk (details in \S\ref{sec:sim}).  This latter model is not intended to be 
a definitive description of these disks, but rather illustrative of the 
broader category of models without a sharp outer edge.  
The model fitting 
process involved deriving a minimum $\chi^2$ solution for those parameters 
of each class of model that best fit the continuum emission, and then using 
standard assumptions to predict CO emission (described in \S\ref{sec:fit}).  
The CO emission was not used to determine the model fits, due to the 
computational intensity of solving the excitation and radiative transfer 
for the molecular line for a large grid of models.

\subsection{Truncated Power Law}
\label{sec:pl}

For the truncated power law models, we used the prescription of \citet{dut94}.
In this framework, the disk structure is described by power laws in temperature
and surface density, with the scale height specified through the assumption 
that the disk is in hydrostatic equilibrium:
\begin{eqnarray}
T(R) = T_{100} \left( \frac{R}{100 AU} \right)^{-q} \\
\Sigma(R) = \Sigma_{100} \left( \frac{R}{100 AU} \right)^{-p} \\
H(R) = \sqrt{ \frac{2 R^3 k_B T_k(R)}{G M_\star m_0}}
\end{eqnarray}
where the subscript `100' refers to the value at 100~AU, $k_B$ is Boltzmann's 
constant, $G$ is the gravitational constant, $M_\star$ is the stellar mass, and
$m_0$ is the mass per particle (we assume 2.37 times the mass of the hydrogen
atom).
Combining these expressions and the assumption of hydrostatic equilibrium, 
the volume density $n(R,z)$ is given by:
\begin{eqnarray}
n(R,z) = \frac{\Sigma(R)}{\sqrt{\pi} H(R)} \exp -(z/H(R))^2
\end{eqnarray}
where $z$ is the vertical height above the midplane.  As noted by 
\citet{dut07}, this definition implies a scale height of 
$H(r) = \sqrt{2} c_s/\Omega$, where $c_s$ is the sound speed and $\Omega$ the 
angular velocity, while other groups use $H(r) = c_s/\Omega$; this difference
should be taken into account when comparing our results with other disk 
structure models.  During the modeling process, we recast the surface density 
normalization in terms of the midplane density at 100 AU, so that the 
parameter $\Sigma_{100}$ is replaced by $n_{100}$.  
This power-law model of disk structure has five free parameters: 
$T_{100}$, $q$, $n_{100}$, $p$, and $R_{out}$.

\subsection{Similarity Solution from Accretion Disk Evolution}
\label{sec:sim}

While versatile and ubiquitous, the truncated power law models of disk 
structure have one obviously unphysical feature: a sharp outer edge.  
In the absence of dynamical effects (e.g. from a binary companion) or 
large pressure gradients to confine the material, disk structure at the 
outer edge is expected to taper off gradually.  A description of the 
structure of an isolated, steadily accreting disk as it evolves with time is 
provided by \citet{har98}, who expand on the work of \citet{lyn74} to show that 
if the viscosity in a disk can be written as a time-independent power law 
of the form $\nu \propto R^\gamma$, then the similarity solution for the disk 
surface density is given by
\begin{eqnarray}
\Sigma(r) = \frac{C}{r^\gamma} T^{-(5/2-\gamma)/(2-\gamma)} \exp \left[ -\frac{r^{2-\gamma}}{T} \right]
\end{eqnarray}
where C is a constant, $r$ is the disk radius in units of the radial scale factor $R_1$ such that $r=R/R_1$, and $T$ is the nondimensional time
$T = t/t_s + 1$ where $t_s$ is the viscous scaling time 
\citep[eq. 20 in][]{har98}.
For simplicity, when applying these models to our data we used physical 
units and absorbed several of the parameters into two constants so that the 
surface density is of the form
\begin{eqnarray}
\Sigma(R)=\frac{c_1}{R^\gamma} \exp \left[ -\left(\frac{R}{c_2}\right)^{2-\gamma} \right],
\end{eqnarray}
where $R$ is the disk radius in AU and $c_1$, $c_2$, and $\gamma$ are constants
that we allowed to vary during the fitting process.

The temperature profile for the similarity solution disk model is identical to 
that of the truncated power-law disk, except that its spatial extent is 
infinite.  We do not allow it to drop below 10 K, but this low temperature 
limit is not problematic for any of the disks considered here.
This model therefore includes five free parameters: $T_{100}$, $q$, $c_1$, 
$\gamma$, and $c_2$.  The constant $c_1$ describes the normalization of the 
surface density, similar to $n_{100}$ in the power-law model, while 
the constant $c_2$ is analogous to the outer radius, since it describes the 
radial scale length over which the exponential taper acts to cause the 
surface density to drop towards zero.  

\subsection{Model Comparison} 
\label{sec:comp}

The surface density description for the similarity solution is similar to 
the truncated power law except at the outer edge of the disk.  
In the inner regions of the disk for which $R \ll c_2$, we may
expand the exponential so that $\exp{\left[-(R/c_2)^{2-\gamma}\right]} \rightarrow 1 - (\frac{R}{c_2})^{2-\gamma} + \cdots$, and the surface density becomes
\begin{eqnarray}
\Sigma(R) = \frac{c_1}{R^\gamma} (1 - \left(\frac{R}{c_2}\right)^{2-\gamma} ) = \frac{c_1}{R^\gamma} - \frac{c_1}{c_2^{2-\gamma}} R^{2(1-\gamma)}
\end{eqnarray}
In the $\alpha$-viscosity context \citep{sha73}, for a vertically isothermal
disk with the typical temperature index $q=0.5$, we would expect that $\gamma=1$.  
This implies that for standard assumptions, the inner disk surface density 
will be described by a power law in $R$ with index $\gamma$, modified by a constant ($\frac{c_1}{c_2}$) due to the influence of the exponential.  
If $\gamma$ deviates from 1, an additional shallow dependence on $R$ would be 
expected.  

It is illuminating to consider the behavior of these models in the Fourier 
domain, the natural space for interferometer observations.
To do so we define the coordinate ${\cal R}_{uv}$, the
distance from the phase center of the disk in the $(u,v)$ plane, as it would
be observed if the disk were viewed directly face-on.  To perform the 
deprojection from the inclined and rotated sky coordinates, we calculate the 
position of each point in the $(u,v)$ plane as a projected distance from the 
major and minor axes of the disk, respectively: $d_a = {\cal R}\sin \phi$ and 
$d_b = {\cal R} \cos \phi \cos i$, where $i$ is the disk inclination, 
${\cal R} = (u^2+v^2)^{1/2}$, $\phi$ is the polar angle from the major 
axis of the disk, $\phi = \arctan (v/u - PA)$, and $PA$ is the position angle 
measured east of north; then ${\cal R}_{uv} = (d_a^2+d_b^2)^{1/2}$
\citep{lay97, hug07}.

In the Fourier domain, the truncated disk models show ``ringing'' 
and the visibilities will drop rapidly to zero in the vicinity of 
${\cal R}_{uv} = 1/R_{out}$.  
Under the simplifying assumption of $\gamma = 1$, the Fourier transform of
the similarity solution becomes a convolution of two functions: 
(1) $1/{\cal R}$, which is just the Fourier transform of the $1/R$ dependence 
of a power law extending from zero to infinity, and (2) 
$c^2/(1+{\cal R}^2c^2)^{3/2}$, where $c$ is a scaling constant for the term 
that describes the exponential taper.  Since these two functions both decrease 
monotonically and are always positive, the visibilities drop smoothly 
to zero without any ringing.

\subsection{Model Fitting}
\label{sec:fit}

For both disk models, we fit for the five parameters describing the disk 
temperature and surface density structure using the continuum data for each 
disk with the widest range of available baseline lengths.  The position 
angle and inclination were fixed and adopted from previous studies 
(see Table~\ref{tab:params}). For opacity, we assume the standard millimeter 
value adopted by \cite{bec91} ($\kappa_\nu = \kappa_0 (\nu/\nu_0)^\beta$, 
where $\kappa_0= 10.0$ cm$^2$/g$_{dust}$, $\nu_0$=1~THz, and $\beta = 1$), 
although we allow $\beta$ to vary in order to obtain the proper normalization 
when extrapolating from one frequency to another.  Due to the $\sim 100$~AU 
spatial resolution, these data are not sensitive to the inner radius of the 
disk.  For both sets of models, therefore, we simply fix the inner radius 
at a value of 4~AU for TW Hya \citep{cal02,hug07}, 24~AU for GM Aur 
\citep{cal05}, and 3~AU for the other two systems, for which reliable inner 
radius information is not available; this is sufficiently small that changes 
in the inner radius do not affect the derived model parameters.  To compare the 
models to the data, we use the Monte-Carlo radiative transfer code RATRAN 
\citep{hog00} to calculate sky-projected images of the dust continuum and 
CO emission, with frequency and bandwidth appropriate for the observations, 
and assuming Keplerian rotation.  We then use the MIRIAD task {\em uvmodel} 
to simulate the SMA observations, with the same antenna positions and 
visibility weights. 

\begin{table*}
\centering
\caption{Stellar and disk properties}
\begin{tabular}{lcccccc}
\hline
 & Spectral & Dist. & Stellar & Age & Disk & Disk \\
System & Type & (pc) & Mass (M$_\sun$) & (Myr) & PA ($^\circ$) & $i$ ($^\circ$) \\
\hline
\hline
HD 163296 & A1V & 122$^a$ & 2.3$^a$ & 5$^b$ & 145$^c$ & 46$^c$ \\
TW Hydrae & K8V & 51$^{d,e}$ & 0.6 & 5-20$^{f,g}$ & -45$^h$ & 7$^h$ \\ 
GM Aurigae & K5V & 140 & 0.8$^i$ & 2-10$^{j,k}$ & 51$^i$ & 56$^i$ \\
MWC 480 & A3 & 140$^m$ & 1.8$^n$ & 7-8$^{n,o}$ & 143$^p$ & 37$^p$ \\
\hline
\end{tabular}
\tablerefs{
($a$) \citet{van98};
($b$) \citet{nat04};
($c$) \citet{ise07};
($d$) \citet{mam05};
($e$) \citet{hof98};
($f$) \citet{kas97};
($g$) \citet{web99};
($h$) \citet{qi04};
($i$) \citet{dut98};
($j$) \citet{bec90};
($k$) \citet{sim95};
($m$) \citet{the94};
($n$) \citet{pie07};
($o$) \citet{sim00};
($p$) \citet{ham06}.
}
\label{tab:params}
\end{table*}

\begin{figure*}
\centering
\includegraphics[width=0.90\textwidth]{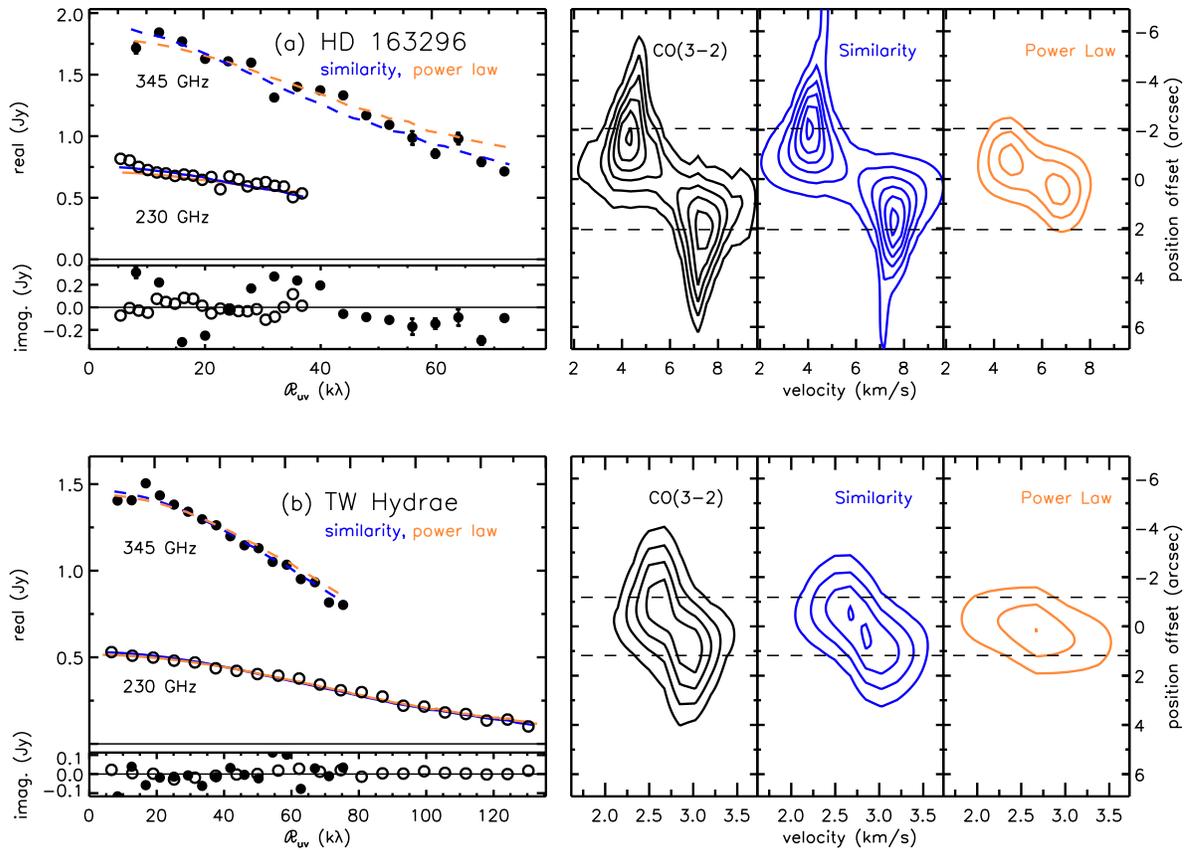}
\caption{
Comparison between the data and the two types of models (similarity solution
and power law) for the four disks
in our sample: {\em (a)} HD 163296, {\em (b)} TW Hydrae, {\em (c)} GM Aurigae,
and {\em (d)} MWC 480.  For each source, the left panel shows the real and 
imaginary visibilities as a function of deprojected $(u,v)$ distance from
the phase center.  
Symbols are SMA data; open circles are 230 GHz and
filled circles are 345 GHz continuum.  The lines represent the best fit to the
345 GHz continuum for the power law (orange) and similarity (blue) models.
Dashed lines show the model at 345 GHz while solid lines are 230 GHz.
The right panel shows position-velocity diagrams of the J=3-2 rotational 
transition of CO along the major axis of the disk.  The left plot (black 
contours) shows the SMA data.  The middle plot (blue contours) displays the 
emission predicted by the similarity solution parameters that provide the 
best fit to the continuum emission, and the right plot (orange contours) 
displays the emission predicted for the best-fit power-law model.  
The horizontal dashed line across the right panel represents the extent of 
the outer radius ($R_{out}$) derived for each source through fitting of the
continuum emission in the context of the power-law model.
The contour levels, beam, and velocity resolution for each source are as 
follows: {\em (a)} [2,4,6,8,10,12]$\times$1.1 Jy/beam, 3.0$\times$2.1
arcsec at a position angle of $14.3^\circ$, and 0.35 km/s; {\em (b)}
[2,4,6,8]$\times$2.0 Jy/beam, 4.0$\times$1.8 arcsec at a position angle of
$3.2^\circ$, and 0.18 km/s; {\em (c)} [2,4,8,12,16]$\times$0.5 
Jy/beam, 2.3$\times$2.1 arcsec at a position angle of $12.9^\circ$, and
0.35 km/s; {\em (d)} [2,4,6,8,10]$\times$0.5 Jy/beam, 
2.5$\times$2.3 arcsec at a position angle of $45.3^\circ$, and 0.35 km/s.
}
\label{fig:bigfig}
\end{figure*}

For each set of parameters, we directly compare the model visibilities to 
the continuum data and calculate a $\chi^2$ value, using the minimum $\chi^2$ 
value to determine the best-fit parameters.  The resulting best-fit models are 
shown along with continuum data for both frequencies in the left panels of 
Figure \ref{fig:bigfig}.  The abscissa gives the deprojected radial distance 
in the $(u,v)$ plane, and the ordinate shows the real and imaginary components 
of the visibility.  For a disk with circular symmetry, the imaginary components 
should average to zero.  The 230 GHz continuum data are depicted as open 
circles, while the 345 GHz data are filled circles.  The best-fit 
power-law model is shown in blue and the similarity solution in orange.  Dotted 
and dashed lines distinguish between the 230 and 345 GHz model predictions; the 
fit was determined at that frequency with the largest baseline coverage and 
extrapolated to the other frequency, by varying $\beta$.  The uncertainties 
quoted for $\beta$ reflect an assumed 10\% calibration uncertainty.  Note that 
varying $\beta$ has no effect on the modeled CO emission.  

\begin{figure*}
\centering
\includegraphics[width=1.0\textwidth]{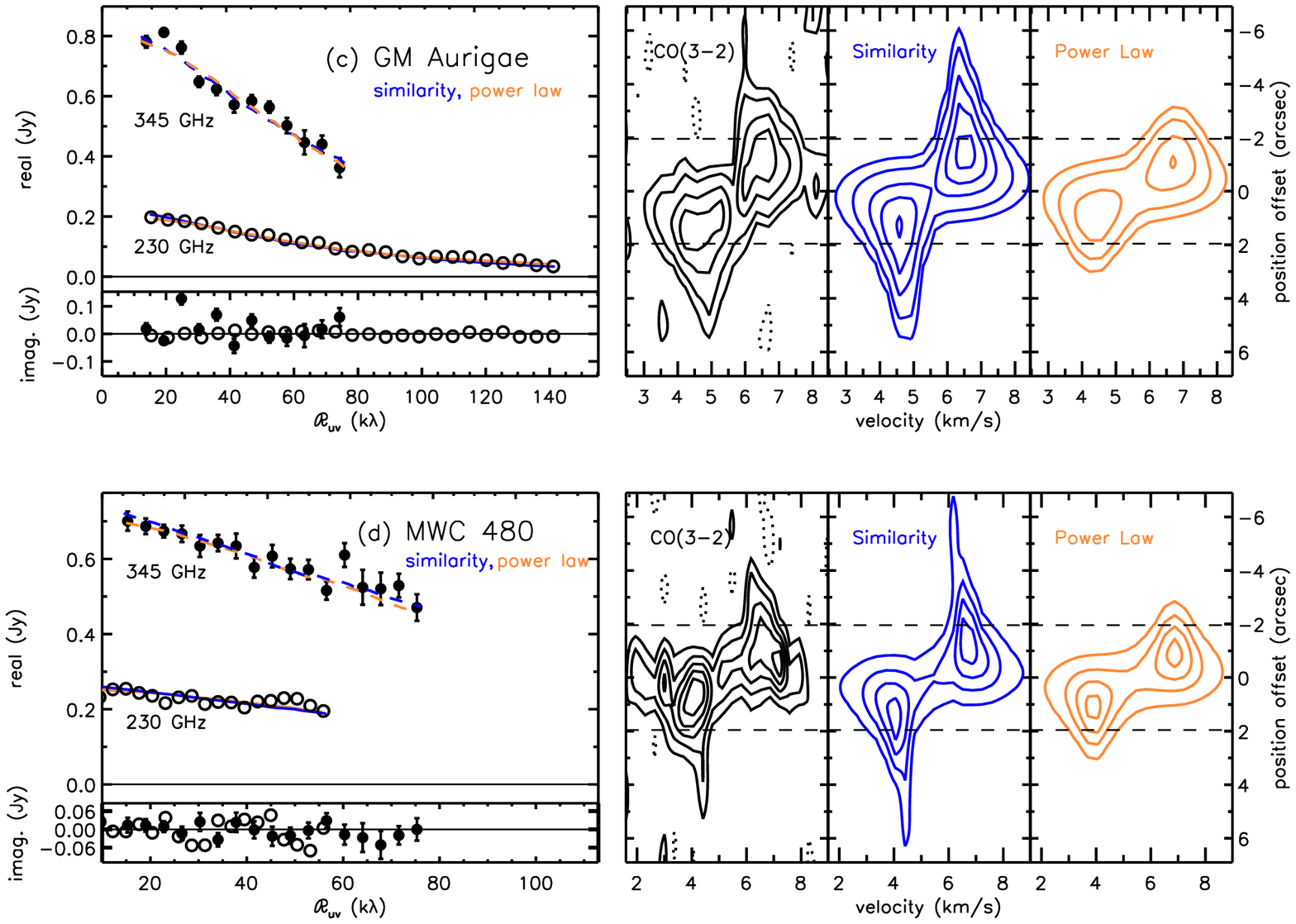}
\end{figure*}

We measure values of $\beta$ that are consistent with 1, which is
in agreement with the typical values measured for disks in the Taurus-Auriga 
association \citep[e.g.][]{dut96,rod06}.  These shallow millimeter spectral 
slopes indicate that some grain growth has occurred from ISM grain sizes, which 
typically exhibit a steeper spectral slope ($\beta \sim 2$).  In particular, 
the value of $1.2$ measured for GM Aur matches well the value 
of $1.2$ reported by \citet{and07beta}, and the value of 
$0.7$ measured for TW Hydrae matches well the value of 0.7 
reported by \citet{cal02} and \citet{nat04}. 

After fitting the continuum, we then assumed a gas/dust mass ratio of 100 and 
a standard interstellar CO/H$_2$ mass ratio of $10^{-4}$ to predict the 
expected strength and spatial extent of CO emission from the disks, based on 
the best-fit continuum model.  We assume throughout that the gas and dust are 
well-mixed, and that CO traces molecular hydrogen. We do not take into 
account the complexities of disk chemistry, such as the depletion of CO 
molecules in the cold, dense midplane \citep{aik07,sem06}.  However, 
deviations from these simple assumptions should have no appreciable effect on 
our conclusions concerning the radial extent of CO emission.

Since we neglect the vertical temperature gradient in the disk, we 
might expect to underpredict the strength of optically thick CO line 
emission, which likely originates in the upper layers of the disk that 
are subject to heating by stellar irradiation. The continuum emission,
by contrast, is likely weighted toward the cold midplane of the disk.
For this reason, after obtaining an initial fit from the continuum,
we allowed the temperature scale ($T_{100}$) to vary to best reproduce 
the flux levels of the observed CO emission, and then iteratively fit for 
the other structural parameters ($q$, $n_{100}$, $c_2/R_{out}$, and $\gamma/p$).

Deriving the temperature from the CO emission in this way may underestimate
the midplane density in some cases, due to the degeneracy between $T_{100}$
and $n_{100}$: the temperature derived from CO emission is typically
greater than or equal to that of the shielded midplane, depending in detail
on the dust opacity and molecular dissociation due to ultraviolet radiation
in the upper disk layers \citep[for a discussion of the processes involved,
see][]{jon07,ise07}.  For the disks considered, the temperature derived for
the dust continuum emission was within $\sim 40$\% of that derived to match
the CO line strength.

\section{Results and Discussion}

\begin{figure}
\includegraphics[totalheight=0.6\textheight]{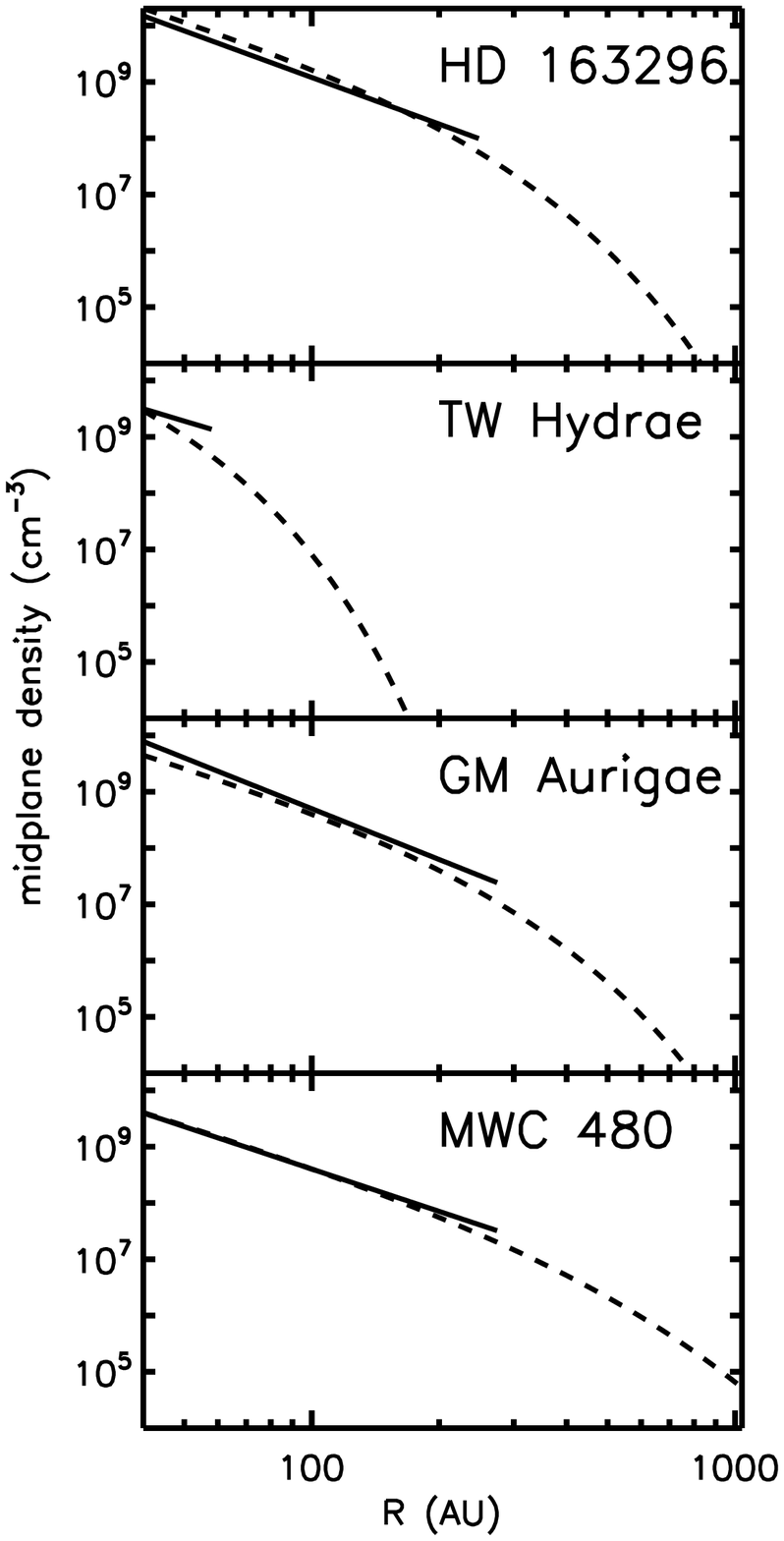}
\caption{
Midplane density structure of the models that provide the best fit to the 
continuum data.  Solid lines show truncated power-law models while dashed 
lines show similarity solution models.  
}
\label{fig:dens}
\end{figure}

The parameters for the best-fit model solutions to the continuum data for 
each source and for each of the two model types are listed in Table 
\ref{tab:fit}.  This table lists only the set of parameters with the minimum 
$\chi^2$ value; formal errors are not quoted as these are not intended to be 
definitive structural models but simply illustrative of the differences 
between the model classes in their treatment of the outer edge.  The midplane 
surface density profiles for these models are plotted in Figure 
\ref{fig:dens}. The solid lines depict the profile for the power-law solution, 
while the dashed lines are for the similarity solution.  The parameters of the 
two model solutions are very similar, particularly for HD 163296 and MWC 480.  
For all four disks, the two model solutions are particularly similar just 
within the outer edge of the disk, around the range of radii well-matched to 
the resolution of the data ($\sim 200$ AU for HD 163296, $\sim 90$ AU for 
TW Hydrae, $\sim 200$ AU for GM Aurigae, and $\sim 300$ AU for MWC 480).  
The outer radius for the power-law solution typically falls at roughly 
twice the scale length ($c_2$) of the similarity solution.
The analogous parameters $\gamma$ and $p$, which describe how 
quickly the midplane density drops with radius, are also very similar between 
the two models. 

\begin{table*}
\centering
\caption{
Parameters for best-fit continuum models  
}
\begin{tabular}{llcccccc|c}
\hline
 & & & & & & $c_2$ (AU) & $\gamma$ & \\
Source & Model & $\chi^2$ & $T_{100}$ (K) & $q$ & $n_{10}$$^a$ (cm$^{-3}$) & $R_{out}$ (AU) & $p$ & $\beta$ \\
\hline
\hline
HD 163296 & Similarity & 2.29 & 65 & 0.4 & $5.3\times 10^{11}$ & 125 & 0.9 & $0.4^{+0.5}_{-0.3}$ \\
 & Power Law & 2.26 & 60 & 0.5 & $6.7\times 10^{11}$ & 250 & 1.0 & $0.5^{+0.5}_{-0.3}$ \\
\hline
TW Hydrae & Similarity & 2.42 & 40 & 0.2 & $2.3\times 10^{11}$ & 30 & 0.7 & $0.7^{+0.5}_{-0.1}$ \\
 & Power Law & 2.41 & 30 & 0.5 & $7.1\times 10^{10}$ & 60 & 1.0 & $0.7^{+0.5}_{-0.1}$ \\
\hline
GM Aurigae & Similarity & 2.19 & 50 & 0.5 & $1.1\times 10^{11}$ & 140 & 0.9 & $1.2^{+0.5}_{-0.1}$ \\
 & Power Law & 2.17 & 40 & 0.4 & $5.0 \times 10^{11}$ & 275 & 1.3 & $1.3^{+0.5}_{-0.1}$ \\
\hline
MWC 480 & Similarity & 1.86 & 50 & 0.8 & $1.0 \times 10^{11}$ & 200 & 1.1 & $0.7^{+0.5}_{-0.4}$ \\
 & Power Law & 1.86 & 45 & 0.7 & $1.3 \times 10^{11}$ & 275 & 1.3 & $0.7^{+0.5}_{-0.4}$ \\
\hline
\end{tabular}
\tablenotetext{a}{Midplane density at 10 AU. We
use the value at 10 AU rather than 100 AU to compare better the power law and
similarity models in the region where their behavior is similar.}
\label{tab:fit}
\end{table*}

CO J=3-2 emission predicted from these best-fit models is shown in the right
panel of Figure \ref{fig:bigfig}.  The similarity solution is shown in the
blue-contoured central plot, and the power-law model in the orange-contoured
plot on the right.  Recessional velocity is plotted on the abscissa while
the position offset along a slice through the disk major axis is plotted on the 
ordinate.  The horizontal dashed line in each figure represents the extent of 
the outer radius ($R_{out}$) derived for that source in the context of the 
truncated power-law model.  For all four sources, the extent of molecular gas 
emission from the similarity solution is much more closely matched to the data 
than that of the power-law model, even though both reproduce the continuum dust
emission equally well.  

From Figure~\ref{fig:bigfig}, it is clear by eye that for all four sources, 
the extent of the CO emission is severely underpredicted by the power law 
model but matches well the predicted emission from the similarity solution 
model.  A calculation of the $\chi^2$ value comparing the predicted CO 
emission for the two models to the observed emission shows that the similarity 
solution matches the data better than the truncated power-law model for all of 
the disks in our study.  The difference is at the $2\sigma$ level for MWC 480, 
for which there is only short-baseline data with relatively low 
signal-to-noise, and at the $4\sigma$ level for GM Aur; for TW Hydrae and 
HD 163296, the $\chi^2$ analysis shows that, formally, the similarity solution 
provides a better fit to the CO emission than the power-law model at the 
$>10\sigma$ level. 

The tapered edge of the similarity solution density distribution evidently 
permits a large enough column density to produce detectable CO 3-2 line 
emission, even though it has dropped off enough that the continuum emission is 
negligible.  The power-law model, by contrast, is strictly limited in the 
extent of its CO emission by the sharp outer radius.  In particular,
for the case of HD 163296, the CO emission predicted by the power law model 
(orange contours in the right panel of Figure~\ref{fig:bigfig}a)
falls to 4.4 Jy/beam at a distance of 1.8 arcsec (220 AU) from the source 
center, while the similarity solution (blue contours) maintains this brightness 
out to a distance of 4.7 arcsec (570 AU).  This latter size is well matched 
to the data (black contours) which extends at this brightness to a distance of 
5.0 arcsec (600 AU).  These distances likely overestimate the true physical 
extent of the disk due to convolution with the $2.1\times3.0$ arcsec beam,
though they are very comparable to the values observed by \citet{ise07}.
While the similarity solution does not provide a perfect fit to the data, nor 
do we intend it to do so, it illustrates that the outer radius discrepancy is 
peculiar to the truncated power-law model; other disk structure models with a 
tapered outer edge may be able to reproduce the gas and dust emission as well 
as, or better than, the similarity solution adopted here.

Analysis of the CO excitation in the similarity solution model shows that 
the extent of the CO J=3-2 emission in these disks coincides roughly with the 
radius at which the line excitation becomes subthermal, determined primarily 
by where the mid-plane density drops below the critical density ($\sim 4.4 
\times 10^4$ cm$^{-3}$ at 20~K, though effectively lowered when photon 
trapping plays a role).  In the similarity solution model, the surface 
density distribution steepens dramatically at large radii, but without the 
sharp truncation of the power-law model.  This suggests that caution should 
be exercised not only when comparing outer radius measurements based on dust 
continuum and molecular gas emission, but also when comparing measurements 
based on emission from different transitions of CO or from isotopologues of 
the CO molecule that have differing abundances and optical depths. 
\citet{pie07} fit truncated power law models to the disks around DM Tau, 
LkCa 15, and MWC 480 in several different isotopologues and rotational 
transitions of CO.  For the two cases in which multiple transitions of the 
$^{13}$CO molecule were observed, the derived outer radius is marginally 
smaller for the J=2-1 transition than the J=1-0 transition.  This result 
is consistent with the expected trend that lower-J transitions will exhibit 
larger outer radii due to their lower critical density: a lower critical 
density will be reached at a greater distance as the surface density tapers 
off near the outer edge of the disk. 
In all cases the \citet{pie07} analysis also yielded 
a smaller outer radius in $^{13}$CO than in $^{12}$CO, as well as a flatter 
surface density power law index for $^{13}$CO than for $^{12}$CO.  
These differences may be related to selective photodissociation, or other
chemical processes. However, the trends of smaller outer radius and shallower 
surface density index in $^{13}$CO are also consistent with surface density 
falling off rapidly at large radii, as expected for a disk with a tapered 
outer edge. In the similarity solution model, the less abundant $^{13}$CO 
isotopologue will become undetectable at smaller radii than $^{12}$CO, which 
is more sensitive to the exponential drop in surface density in the outer disk.

It is noteworthy that studies of six largest ``proplyds'' with the most 
distinct silhouettes in the Orion Nebula Cluster reveal radial profiles 
in extinction that are well-described by an exponential taper at the outer 
edge \citep{mcc96}. These isolated disks may be analogous to the systems
considered here with a tapered outer edge.

Models with tapered outer edges also aid in addressing discrepancies
between the size of the dust disk observed in the millimeter and the
extent of scattered light observed in the optical and near-infrared.
For example, coronographic observations of TW Hydrae detect scattered light
to a distance of $\sim 200$~AU from the star \citep{kri00,tri01,wei02},
while the truncated power-law model places the outer edge of the dust disk
closer to 60~AU.  Similarly, observations of HD 163296 by \citet{gra00}
detect scattered light out to $\sim 400$ AU from the star, much larger than
the 250~AU radius of the dust disk implied by the truncated power-law
model.	While the exponential taper causes the density of the similarity
solution to drop rapidly with radius, these models retain a substantial
vertical column density for several exponential scale lengths. It is
therefore plausible that scattered light can remain visible at this distance,
in contrast to the case of the smaller truncated power-law disk.

Although we intend for the similarity solution applied here to be an
illustrative rather than definitive description of the disk structure, it
is important to note that the particular form applied here has potential
implications for the study of the evolutionary status of these disks.
The form of the similarity solution developed by \citet{lyn74} and
\citet{har98} relates the observed structure to the disk age, viscosity,
and initial radius.  Although all three of these variables are poorly
constrained by current observations, a large and homogeneous sample of
objects studied in this way might reveal evolutionary trends in the disk
structure.

\section{Summary and Conclusions}

With the advent of high signal-to-noise interferometer observations that
resolve the outer regions of nearby protoplanetary disks, an apparent 
discrepancy has emerged between the extent of the dust continuum and 
molecular gas emission \citep{pie05,ise07}. Using multi-frequency 
interferometric data from the Submillimeter Array, we have investigated this 
disparity for four disk systems (HD 163296, TW Hydrae, GM Aurigae, and MWC 480) 
in the context of two distinct classes of disk structure models: (1) a 
truncated power law, and (2) a similarity solution for the time evolution of 
an accretion disk. The primary difference between these models is in their 
treatment of the disk outer edge: the abruptly truncated outer edge of the 
power-law disk causes the visibilities to drop rapidly to zero, leading to an 
inferred outer radius that is small in comparison with the observed molecular 
gas emission.  The similarity solution, by contrast, tapers off smoothly, 
creating a broader visibility function and allowing molecular gas emission 
to persist at radii well beyond the region in the disk where continuum 
falls below the detection threshold.  The outer radius discrepancy appears to 
exist only in the context of the power-law models.  

In light of this result, it appears that an abrupt change in dust properties 
for these disks is unlikely, as there is no physical mechanism to explain such 
a discontinuity. This may imply that a sharp change in dust properties in the 
early solar nebula is similarly an unlikely explanation for the Kuiper belt 
edge observed by \citet{jew98}, and that a dynamical mechanism such as 
truncation by a close encounter with a cluster member \citep[][and references 
therein]{rei05} may provide a more plausible origin.  In this case, we would 
expect to observe disks with sharp outer edges only in clustered environments,
and a model with a tapered edge would be a more realistic prescription for 
investigating the structure of a typical isolated disk.  The tapered disk 
models provide a natural explanation for the disparate outer radii observed 
using different probes of the disk extent, including comparison of continuum 
and molecular gas observations \citep{pie05,ise07}, and also comparison of 
different isotopologues and rotational transitions of a particular 
molecule \citep{pie07}.  When predicting CO emission, this simple model does 
neglect potential variance in the CO abundance due to depletion in the 
midplane and photodissociation at the disk surface; however, the results 
presented are intended simply to illustrate the global differences between 
gas and dust emission from the two model classes, independent of detailed 
CO chemistry.

While we cannot rule out disparate gas and dust radii in these disks, we show 
that an alternative disk structure model, grounded in the physics of accretion, 
resolves the apparent size discrepancy without the need to invoke dramatic 
changes in dust opacity, dust density, or dust-to-gas ratio in the outer disk. 

\acknowledgements
The authors would like to thank Sean Andrews for thoughtful comments that 
helped to improve the manuscript. Partial support for this work was provided 
by NASA Origins of Solar Systems Program Grant NAG5-11777.  A.~M.~H. 
acknowledges support from a National Science Foundation Graduate Research 
Fellowship.

\end{document}